\documentclass[12pt]{iopart}

\usepackage{iopams}

\usepackage{graphicx}

\begin{document}

\title[]{Large-scale defect accumulations in Czochralski-grown silicon}

\author{V.~P.~Kalinushkin\footnote{E-mail: VKALIN@KAPELLA.GPI.RU.},
A.~N.~Buzynin, V.~A.~Yuryev\footnote{E-mail: VYURYEV@KAPELLA.GPI.RU.},
O.~V.~Astafiev\footnote{E-mail: ASTF@KAPELLA.GPI.RU.}}

\address{General Physics Institute of the Russian Academy of Sciences,
38, Vavilov Street, Moscow, GSP--1, 117942, Russia}

\begin{abstract}
Czochralski-grown silicon crystals were studied by the techniques
of the low-angle mid-IR-light scattering and electron-beam-induced current.
The large-scale accumulations of electrically-active impurities detected in
this material were found to be different in their nature and formation
mechanisms from the well-known impurity clouds in a FZ-grown silicon.
A classification of the large-scale impurity accumulations in CZ Si
is made and point centers constituting them are analyzed in this paper.
A model of the large-scale impurity accumulations in CZ-grown Si is also
proposed.
In addition, the images of the large-scale impurity accumulations
obtained by means of the scanning mid-IR-laser microscopy are demonstrated.
\end{abstract}

\section{Introduction}
The detection of the large-scale impurity accumulations (LSIAs) with
the sizes ranged from several to several tens $\mu$m in CZ Si by means
of the low-angle light scatter (LALS) \cite{4}
was reported for the first time in
Ref.\,\cite{1}. It was supposed in that work that LSIAs are analogous in their
nature to the oxygen and carbon clouds observed in FZ Si in Ref.\,\cite{2}.
It was shown, however, as a result of the research of Si crystals grown at
variable growth rate done by LALS and EBIC that most of LSIAs in CZ Si
have a shape close to cylindrical \cite{3} which contradict the cloud
model \cite{1}. In the present work, an attempt is made to select different
in their nature types of LSIAs in CZ Si and an information about their
parameters as well as the influence of different thermal treatments on them
is given.

\section{Experimental details}
Industrial substrates of CZ~Si:B studied in this work
were grown in the $<$100$>$
and $<$111$>$ directions and had the specific resistivity from
1 to 40~$\Omega$\,cm. The specific resistivity of wafers of CZ~Si:P,
which were also investigated in the work, were from  2 to 20~$\Omega$\,cm.
The oxygen concentration in the material ranged from
6\,$\times$\,10$^{17}$ to 10$^{18}$~cm$^{-3}$, the carbon concentration was less
than 10$^{16}$\,cm$^{-3}$.

The investigation was carried out by LALS and EBIC. CO- and CO$_2$-lasers
oscillating at the wavelength of 5.4 and 10.6~$\mu$m, respectively,
were used in LALS to select the scattering by free carrier accumulations
\cite{5}. To determine the activation energies ($\Delta E$) of the centers
constituting LSIAs, the temperature dependances of LALS
intensity were investigated
in the range from 85 to 300\,K \cite{6}. A shape of LSIAs was determined from the
dependances of the LALS diagrams on the sample orientation with respect to
the detection plane. The plasma etching of the sample surface
in a special regime before the Schottky barrier creation greatly increased the
sensitivity of EBIC to electrically-active defects in crystals \cite{7}.

Besides, the images of LSIAs were obtained by means of the scanning
mid-IR-laser microscopy in both modes: scanning LALS (SLALS) and
optical-beam-induced LALS (OLALS) \cite{7a}.
The 10.6-$\mu$m emission of CO$_2$-laser
was used as a probe beam, the 0.97-$\mu$m laser radiation served for
excess carrier generation in OLALS.

In the experiments on annealings, wafers were cut into four sections. One
of them was not treated, the others were subjected to either isothermal
\begin{figure}[t] 
\begin{center}
\includegraphics[scale=1.4]{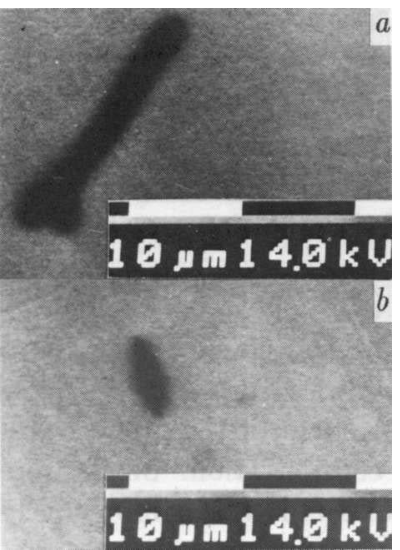}
\includegraphics[scale=0.913]{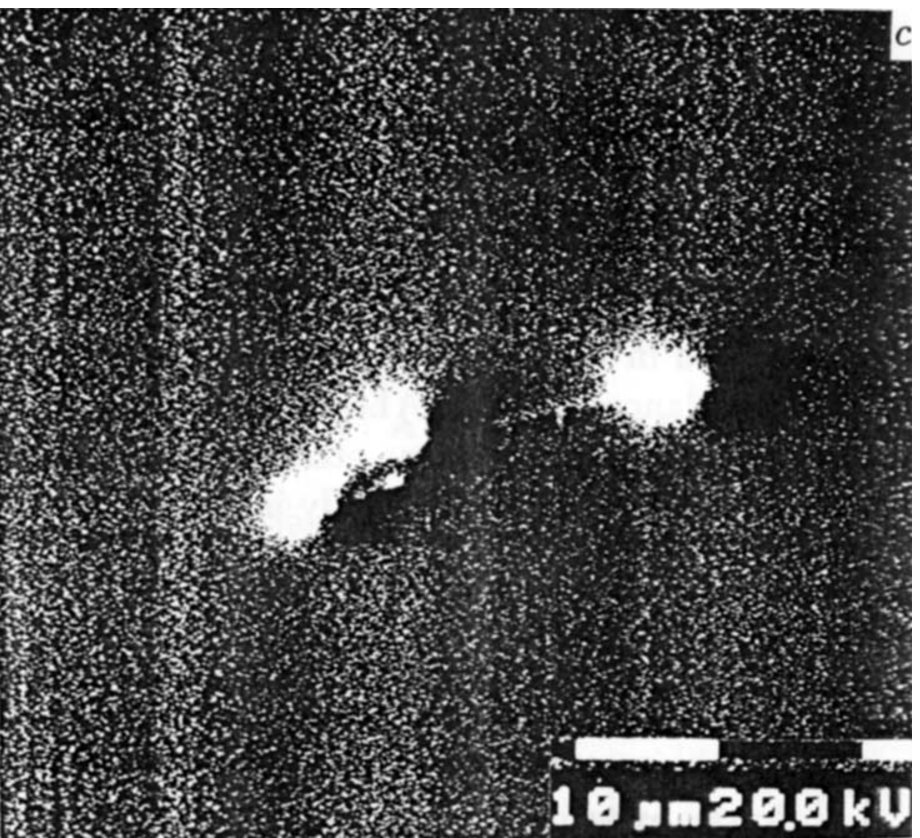}\\
\includegraphics[scale=.927]{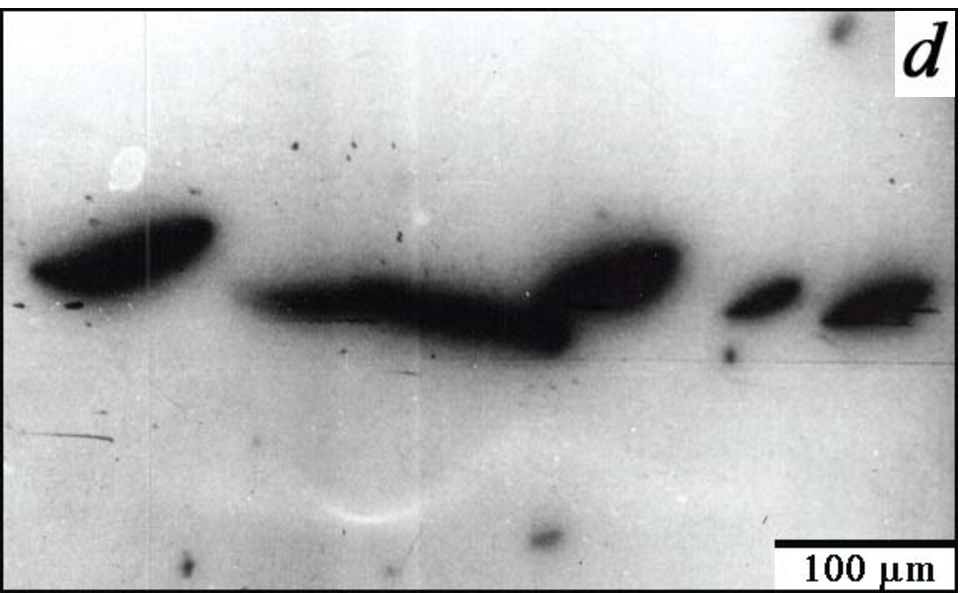}
\includegraphics[scale=.97]{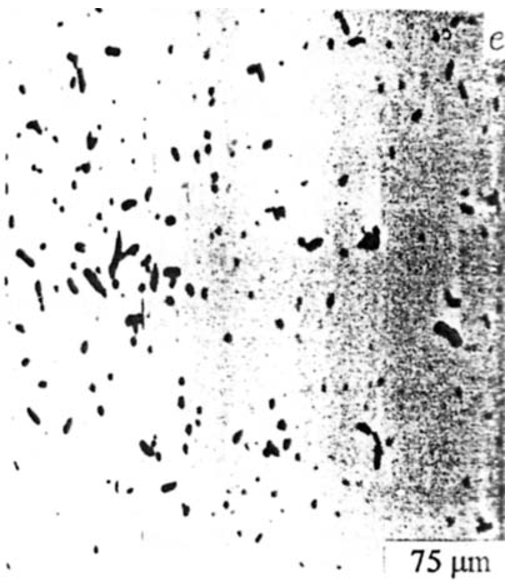}
\end{center}
\caption{EBIC microphotographs of as-grown CZ Si:B:
cylindrical $(a,b)$, spherical $(c)$ and superlarge $(d)$
defects, and a picture showing a distribution of
defects $(e)$.}\label{f1}
\end{figure}
processes at 600 or 800$^{\circ}$C for 24, 48 and 120~h, respectively, or
high-temperature treatments at 965, 1100, 1150, 1200 and 1250$^{\circ}$C
for several tens minutes. The treatments of CZ~Si:B
at $T > 1200^{\circ}$C resulted in
the formation of a large amount of defects of structure which were
revealed by the selective etching (SE). The substrates grown in the $<$100$>$
direction were subjected to both the former and the latter treatments, while
those grown in the $<$111$>$ direction were treated only in the latter way.

\section{CZ Si:B}
\subsection{Initial samples}

Fig.\,\ref{f1} demonstrates the EBIC images of defects. The samples
shown there
contain many non-uniformities with the sizes from
several to several tens $\mu$m\,---\,mainly cylindrical. In
addition, spherical defects are also seen. We have classified
the observed defects in the following way.

\subsubsection{Cylindrical defects {\rm (CDs)}.}
There are sections of LALS diagrams, the shape of
which is dependent on the sample orientation to the detection plane
(Fig.\,\ref{f2}, $\theta < 4.5^{\circ}$). These sections are well
fitted with the curves of scattering by cylinders \cite{4} with the diameters
from 3--4 to 8--10~$\mu$m and length from 15 to 40~$\mu$m depending on a
sample. They predominantly oriented along the $<$110$>$ direction.
It is seen from the EBIC patterns that CDs have rather elliptical or
curved-cylindrical shape (Fig.\,\ref{f1}\,$(a,b)$). We assume the CDs
revealed by EBIC and LALS to be the same defects
similar to CDs observed in Ref.\,\cite{3}.

The concentration of CDs\,---\,the most usual defects in CZ Si:B\,---\,estimated
from EBIC ranged from 10$^6$ to 10$^7$~cm$^{-3}$. We could not find
a dependance of the CD concentration on oxygen concentration, growth
direction, ingot diameter or location on a wafer. Nonetheless we found
their concentration to vary within a wafer as well as in different wafers.

LALS measurements at 10.6 and 5.4-$\mu$m wavelength showed
CDs to be domains of the enhanced
free carrier concentration \cite{5}. Using the CD concentrations from
EBIC, the variations of the dielectric function ($\Delta \varepsilon$) and
the maximum free carrier concentration in them  ($\Delta n_{max}$) were
evaluated \cite{4}: they are (1--4)\,$\times$\,10$^{-4}$ and
(3--10)\,$\times$\,10$^{15}$\,cm$^{-3}$, respectively.
CDs occupy less than 3\,\% of crystal volume and the total amount of
impurities contained  in them ($N_i$) is not greater than
3\,$\times$\,10$^{14}$\,cm$^{-3}$.

LALS temperature dependances showed for CDs a small (2 to 3 times)
drop of the scattering intensity ($I_{sc}$) at about 90\,K
(Fig.\,\ref{f3}, curve 1). This allows us to state that LALS by CDs
at 300\,K at 10.6\,$\mu$m is controlled by the centers with
$\Delta E \approx$\,40--60\,meV containing in CDs. Naturally
another defects, such as deep or compensating centers as well as precipitates,
inclusions and  structural imperfections can also be contained in CDs.

\subsubsection{Spherical defects {\rm (SDs)}.} 
\begin{figure}[th]
\begin{center}
\includegraphics[scale=2.8]{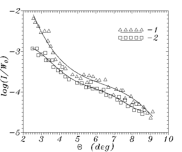}
\includegraphics[scale=2.8]{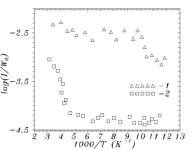}
\includegraphics[scale=2.8]{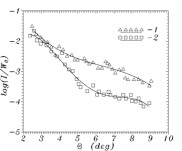}
\end{center}
\caption{LALS diagrams for the as-grown CZ Si:B, orientation
with respect to the detection plane (deg.): 0 (1), 90 (2).}\label{f2}
\caption{Dependances of LALS intensity on
sample temperature  for the cylindrical (1)
and spherical (2) defects in the as-grown CZ Si:B.}\label{f3}
\caption{LALS diagrams at different temperatures (K):  300 (1), 110
(2).}\label{f4}
\end{figure}

Besides the above sections of
LALS diagrams, those independent of the sample orientation were also observed
(Fig.\,\ref{f2}, $\theta > 4.5^{\circ}$). These sections are well
fitted with the curves of light scattering by spherical defects with
the Gaussian profile of $\varepsilon$ \cite{4}
and sizes from 5--8 to 20~$\mu$m. Such defects are also seen in the EBIC
pictures (Fig.\,\ref{f1}\,$(c)$). Their concentration is usually about
10$^5$\,cm$^{-3}$, $\Delta \varepsilon \approx$\ (1--3)\,$\times$\,10$^{-3}$,
$\Delta n_{max} \approx$\ (3--9)\,$\times$\,10$^{16}$\,cm$^{-3}$ \cite{4}.
SDs occupy less than 0.04\,\% of the crystal volume,
$N_i\   _{^{\sim}}^{_<}\ 4\,\times\,10^{13}$\,cm$^{-3}$.

LALS temperature dependances showed SDs, like CDs, to be domains with
enhanced concentration of ionized at 300\,K impurities with
$\Delta E \approx$\,120--160\,meV. These impurities are ``frozen out'' at
about 250\,K, so SD-related scattering is smothered in the range from
90 to 250\,K which enables the accurate selection of CD-related scattering in
LALS diagrams (Fig.\,\ref{f4}).

\subsubsection{Superlarge {\rm (SLDs)} and small {\rm (SmDs)} defects.} The
sections of LALS diagrams well approximated with the curves of light scattering
by defects with the sizes greater than 50\,$\mu$m were sometimes observed.
SLDs were sometimes seen in the EBIC photographs (Fig.\,\ref{f1}\,$(d)$).
These defects appeared to have an asymmetrical shape.

The sections of LALS diagrams independent of the scattering angle
(``plateaux'') were often observed at 5.4\,$\mu$m (and sometimes at
10.6\,$\mu$m) which correspond to defects with the sizes less than
4--5\,$\mu$m. $I_{sc}$ for SmDs was also independent of the probe
wavelength, so they are  domains with the enhanced free carrier concentration.
SmDs were sometimes observed in the EBIC picture as well. Although we
could not unambiguously determine their shape, SmDs seem to be very small
CDs and SDs rather than a separate class of defects. This was verified
\begin{figure}[t]
 \begin{center}
\includegraphics[scale=1.5]{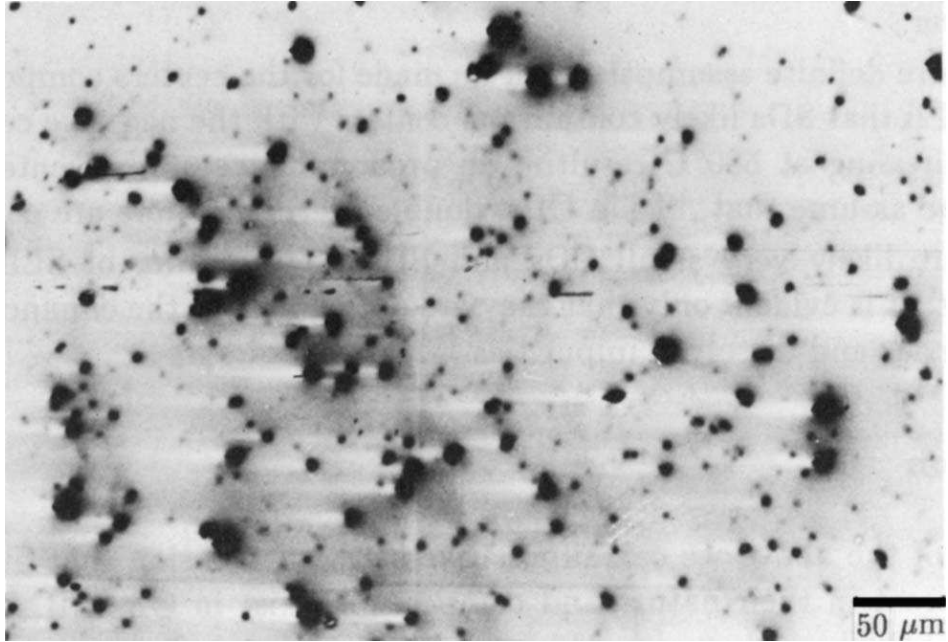}
\end{center}
\caption{Typical EBIC microphotograph of defects in CZ Si:B annealed
at $T\,>\,1100^{\circ}$C.}\label{f5}
\end{figure}
by LALS temperature dependances: SmDs were ``frozen out'' at 90\,K
when CDs predominated and at 250\,K if SDs predominated.

\subsection{Annealed samples}
The LALS diagrams and EBIC pictures for the annealed crystals did not
differ in general features from those for the as-grown samples. The
following peculiarities may be emphasized.

1. In crystals annealed in the temperature range of 600--1100$^{\circ}$C,
the light scatter by SDs was greatly (but not completely) suppressed,
and CDs and SmDs predominated. EBIC showed mainly CDs and SmDs
too.

2. Annealing at $T > 1100^{\circ}$C resulted in predominance of
SD-related scattering and general growth of $I_{sc}$. A great number
of SDs was observed by EBIC (Fig.\,\ref{f5}).

3. After annealing at $T > 1200^{\circ}$C, the centers with the same $\Delta E$
as in the as-grown samples composed LSIAs ($\Delta E \approx$\,120--160\,meV).

4. Annealing at 800$^{\circ}$C and short (up to 48\,h) treatment at  600$^{\circ}$C
did not change $\Delta E$ of the centers composing CDs and SDs. Longer
treatment at 600$^{\circ}$C resulted in prevailing of the centers with
$\Delta E \approx$\,70--90\,meV in CDs and SDs.

5. After 120-h annealing at 600 and 800$^{\circ}$C, SLDs became more habitual
than in the as-grown samples. The centers with $\Delta
E \approx$\,130--170\,meV
were contained in SLDs.

\section{CZ Si:P}

The data for CZ Si:P are much less complete than those for CZ~Si:B.
This material was investigated only by LALS. The main results for CZ~Si:P
can be summarized as follows.

1. The mid-IR-light scattering intensity by the as-grown samples
was as a rule rather low, it was in excess of 2 orders of magnitude weaker than
that for CZ~Si:B\,\footnote{The similar regularity was observed for FZ~Si in
passing from $p$-type to $n$-type material \cite{1}.}. Usual shape of the LALS
diagrams was ``plateau'' which corresponds to the scattering by defects with the
sizes less than 4\,$\mu$m, although the diagrams typical for the CZ~Si:B
crystals reported above were observed sometimes too. These crystals showed the
standard for CZ~Si:B mid-IR-light scattering intensity and set of defects.

2. Annealings at moderate temperatures ($T > 600^{\circ}$C) resulted in
manifestation of the same set of defects that is characteristic for
the CZ~Si:B crystals. General growth of the mid-IR-light scattering
intensity was also observed\,\footnote{It should be mentioned
that $I_{sc}$ as well as the defect size $a$ as a function of the annealing
temperature is very non-monotonic and consists of a set of maxima and
minima \cite{2}. This behaviour is common for both boron and phosphorus
doped materials and very different from that in FZ~Si where routine
diffusion process takes place \cite{1}.}.

\section{SLALS and OLALS images}

Recently, a technique of the scanning mid-IR-laser microscopy has been
proposed by us for LSDA and LSRD visualization in semiconductors \cite{7a}.
As it is
based on LALS, we use the titles ``scanning LALS (SLALS)'' and
``optical-beam-induced LALS (OLALS)'' for its two main modes which are
designed for the investigation of LSDAs and LSRDs, respectively.

\begin{figure}[t]
 \begin{center}
\includegraphics[scale=.95]{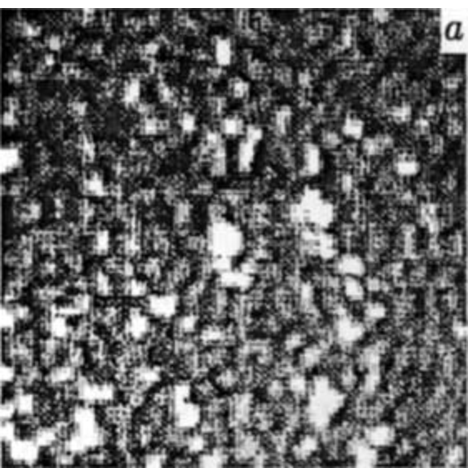}
\includegraphics[scale=.8]{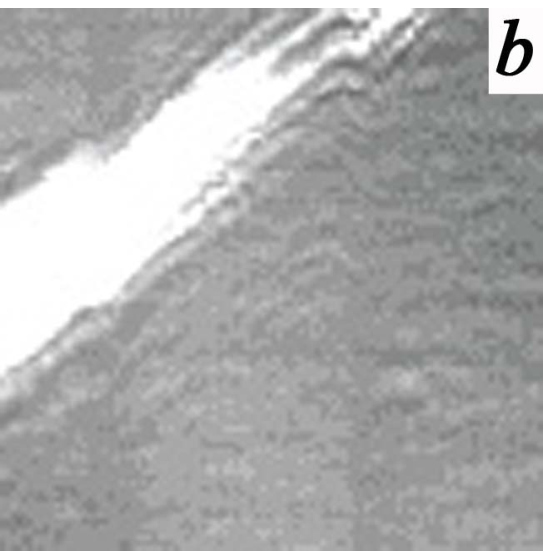}
\includegraphics[scale=.8]{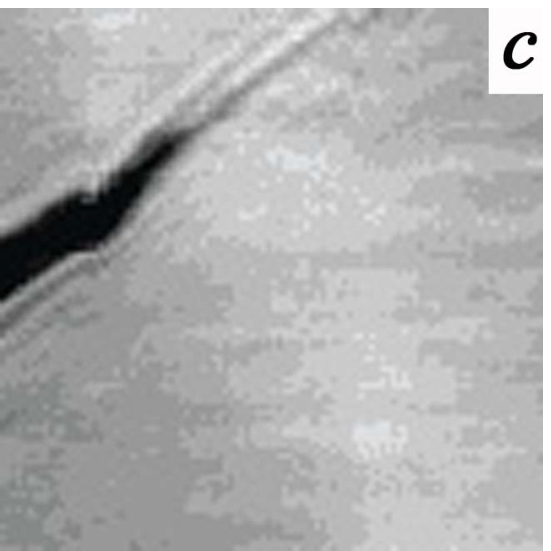}
\end{center}
\caption{SLALS $(a,b)$ and OLALS $(c)$ images of defects in
CZ~Si: defects in the as-grown CZ~Si:P ($\varrho =4.5\,\Omega$\,cm) $(a)$;
superlarge defects (SLDs) in CZ~Si:B
($\varrho =20\,\Omega$\,cm), boundary of $A$ and $A^{\prime}$
zones $(b,c)$; the lighter SLALS image the higher scattering
intensity is, the darker OLALS image the less excess carrier
lifetime; the areas of 1\,$\times$\,1~cm$^2$ are depicted.}\label{SLALS}
\end{figure}

The measurements made by this technique have shown a good agreement
with our former observations made by EBIC, so the estimates of
the parameters of defects made using EBIC and given above
were confirmed by the direct method. Fig.\,\ref{SLALS} demonstrates
pictures of defects in CZ~Si crystals measured by SLALS and OLALS.
A great number of LSDAs were observed in CZ~Si:P with
$\varrho =4.5\,\Omega$\,cm which was related to the
group of samples with high $I_{sc}$ and a set of defects
registered by LALS analogous to that in
CZ~Si:B (Fig.\,\ref{SLALS}\,$(a)$, white spots are the images
of defects).  Unfortunately, now we cannot judge about the defects sizes and
shapes from SLALS images, as the resolution of the instrument we use at present
is two bad \cite{7a}, but at least the defect concentrations can be evaluated.

Fig.\,\ref{SLALS}\,$(b,c)$ demonstrates the
SLALS and OLALS images
of a defect which looks like SLD (or a group of SLDs) and is very similar to
the image of a group of SLDs shown in Fig.\,\ref{f1}\,$(d)$. This
picture was obtained from the boundary region between $A$ and $A^{\prime}$
zones in CZ~Si:B crystal with $\varrho =20\,\Omega$\,cm, i.e. in the
conditions analogous to those in which SLDs usually were observed.

\section{Discussion}

It is difficult now to determine the nature of LSIAs in CZ Si:B unambiguously.
Ii is clear, however, that CDs are domains with the enhanced free carrier
concentration caused by point centers with $\Delta E \approx$\,40--60\,meV.
CDs look like the defects observed by EBIC after oxidizing annealings
\cite{8}. Our research gave an evidence to the presence of these defects
in initial crystals, moreover the CD-related sections of LALS diagrams
do not change after high-temperature annealings. In our opinion, however,
CDs are the impurity atmospheres (IAs) around some defects-precursors, e.g.
stacking faults (SFs). Remark that some authors connect the contrast of
EBIC patterns with the formation of precipitate colonies around SFs \cite{8}.
As these colonies may have no influence on the free carrier concentration
in IAs, the following scenario might be proposed.

In initial wafers, the precipitate concentration in IAs is low.
At the same time, the dissolved
ionized impurity concentration  is high enough but insufficient for
the precipitate formation, hence $I_{sc}$ is high and the recombination
contrast in EBIC is  low. During high-temperature annealing,
precipitate colonies arise in CDs but the dissolved impurity ($N_i$)
and free carrier ($\Delta n$) concentrations
change weakly (e.g. because the impurity
concentration in CDs in the as-grown samples was close to the saturation limit
or due to the growth of the compensation degree). The recombination contrast
will have grown and $I_{sc}$ will have changed weakly and randomly.

From the other hand, the enhanced EBIC contrast may be caused by the
specificity of the sample preparation. Two variants are possible.
The centers enhancing the EBIC contrast may arise as a result of
the plasma etching applied. Alternatively, an ``exhaustion'' of CDs may be
a result of the chemical etching usually applied for sample preparation
for EBIC.

Thus, the hypothesis according to which CDs are IAs around SFs does not meet
contradictions. As to the point defects composing CDs, we suppose them to be
the ``new'' thermal donors \cite{9} whose
activation energies are close to the estimates
made\,\footnote{Although, some alternatives exist \cite{10}, and B and [Cu--O]
are among them.}. The influence of annealing
at 600$^{\circ}$C on the activation energy $\Delta E$
of the centers composing CDs indirectly verifies the assumption\,\footnote{The
growth of ionization energy of ``new'' thermal donors as a result of long-term annealing at 650$^{\circ}$C was reported in Ref.\,\cite{11}.}. New
experiments are required to obtain more evidences to the model
proposed, though.

SDs also are domains with the enhanced dissolved impurity and free carrier
concentrations. We assume SDs to be IAs around defects of structure (e.g.
precipitates). This hypothesis is confirmed by the growth of the SD
concentration during the high-temperature annealings and correlation
with the appearance of structural defects revealed by SE. This assumption
also have no sufficient evidences\,\footnote[1]{Some alternatives
to the assumption\,---\,the cloud models\,---\,are discussed in
Ref.\,\cite{10}.} and require an additional research, however.

Some more definite assumption can be made for the centers composing SDs.
It was supposed in Ref.\,\cite{12} that SDs likely contain the centers with
the negative correlation energy. Regarding annealing at
600$^{\circ}$C resulting in
predominance of the centers with changed $\Delta E$ in SDs, we assume
that, like in CDs, double thermal donors are contained
in SDs\,\footnote{Possible alternatives to these centers are discussed in Ref.\,\cite{10}, they are
B$_i$, [O--V], [C$_i$--C$_s$], {\em etc}.}.

SmDs are likely very small CDs and SDs. The nature of SLDs is hard to be
discussed now. It is evident only that they are domains with the enhanced
concentration of the free carrier and dissolved impurities.

\section{Conclusion}

On the basis of the above we can summarize in the conclusion that
at least two types of LSIAs different in their nature and composition exist in
CZ Si:B. Their parameters determined for the group of crystals
investigated in this work
are rather typical for the industrial Si:B with the specific
resistivity from several to several tens $\Omega$\,cm.

\ack
This work was partially financed by the Russian
Foundation for Basic Researches (grant No.~96-02-19540). The authors
acknowledge RFBR for the support.\\

\newcommand{\newref}{\\\hspace*{-10pt}}

\clearpage

\oddsidemargin  0.25in
\evensidemargin 0.25in
\textwidth 160mm
\textheight 240mm
\pagestyle{empty}


\begin{center}
\Large
\bf
LARGE-SCALE DEFECT ACCUMULATIONS
IN CZOCHRALSKI GROWN SILICON\\[3ex]

V.~P.~Kalinushkin, A.~N.~Buzynin, V.~A.~Yuryev, and O.~V.~Astafiev\\[2ex]

\normalsize
\sl
General Physics Institute of RAS, 38, Vavilov Street, Moscow,
117942, GSP--1, Russia\\[1ex]

Tel./Fax: 7 (095) 135 13 30 \\[0.5ex]
E-mail: vkalin@ldpm.gpi.ru\\[3ex]
\end{center}
\rm

It was reported in {\sl Ref.\,[1]} about detecting of the large-scale
electrically
active defect accumulations (LSDAs) in CZ Si. In this work,  the  data
on their nature are presented.

   The method of low-angle light  scattering  (LALS)${\sl ^2}$  including
scanning LALS (SLALS)$\sl {^3}$, EBIC and selective etching have been used
in the experiments. The crystals grown  under  different  conditions
and subjected to different treatments have been studied.

   It has been shown that CZ Si crystals contain the following types
of LSDAs:
\begin{sloppypar}
\begin{enumerate}
\item
{\em Cylindrical defects} range by diameters from 3--4 to 8--10\,$\mu$m and
by lengths from 10 to 40$\,\mu$m. They line up with crystallographic
directions. Their concentration makes up to 10$^7\,$cm$^{-3}$,
the free carrier
concentration in them is (1--10)$\times$10$^{15}$\,cm$^{-3}$. The centers with
activation energy of 40--60\,meV\,---\,presumably ``new  thermal
donors''---\,form them.
\item
{\em Spherical defects} have the radii up to 15\,$\mu$m. Their concentration
ranges within 10$^4$--10$^5$\,cm$^{-3}$. They are composed by  point  centers
with activation energy of 130--170\,meV. These centers are likely
metastable ones with negative correlation  energy,  probably,
oxygen--vacancy complexes and double thermal donors. The free  carrier
concentration in these LSDAs is (5--50)$\times$10$^{15}$\,cm$^{-3}$.
\item
{\em Superlarge defects} are those of intricate shape with the sizes
greater than 50\,$\mu$m. They have been observed most  often  in  silicon
obtained in ``vacancy-rich'' regime of growth. The nature of these
defects has not been found out yet.
\item
The crystals of CZ Si contain also a  large  number  of  {\em small
defects} with the sizes less than 2--3\,$\mu$m. At present, their  detailed
investigation is difficult because of limitations  imposed  on  LALS
and EBIC. It may be supposed, however, that at least in a number  of
cases they are spherical or cylindrical defects of very small dimensions.
\end{enumerate}
\end{sloppypar}
\begin{flushleft}
$\sl ^1\,$A.~N.~Buzynin {\em et al, Sov.\,Phys.--Semicond.}
{\bf 24} (1990) No~2.\\
$\sl ^2\,$V.~P.~Kalinushkin, {\em Proc.\,Inst.\,Gen.\,Phys.\,Acad.\,Sci.\,USSR},
Vol.\,4,  New York, Nova, 1988, p.1.\\
$\sl ^3\,$O.~V.~Astafiev {\em et al, Proc. SPIE} {\bf 2332} (1994) 138.\\
\end{flushleft}

\end{document}